\documentclass[prb, twocolumn,amsmath,amssymb, superscriptaddress,nofootinbib]{revtex4-1}
\usepackage{graphicx}% Include figure files
\usepackage{natbib,hyperref,soul}
\usepackage{amsmath,epsfig,color,amssymb}
\usepackage[normalem]{ulem}

  \definecolor{new}{rgb}{.08,.05,.8}

\newcommand{\arrd}{\textcolor{black}}

\begin{document}

\author{Areg Ghazaryan}
\affiliation{IST Austria (Institute of Science and Technology Austria), Am Campus 1, 3400 Klosterneuburg, Austria}

\author{Emilian M. Nica}
\affiliation{Department of Physics, Arizona State University Tempe, Arizona 85287-1504, US}

\author{Onur Erten}
\affiliation{Department of Physics, Arizona State University Tempe, Arizona 85287-1504, US}

\author{Pouyan Ghaemi}
\affiliation{Physics Department, City College of the City University of New York, New York, NY 10031, USA}
\affiliation{Physics Program, Graduate Center of the City University of New York, NY 10031, USA}

\title{Shadow surface states in topological Kondo insulators}

\begin{abstract}
The surface states of 3D topological insulators 
in general have negligible quantum oscillations 
when the chemical potential is tuned to the Dirac points. 
In contrast, we find that topological Kondo insulators 
can support surface states with an
arbitrarily large Fermi surface
when the chemical potential is pinned to the Dirac point. 
We illustrate that these Fermi surfaces 
give rise to finite-frequency quantum oscillations, which can become comparable to the extremal area of the unhybridized bulk bands. We show that
this occurs when the crystal symmetry is lowered from cubic to tetragonal in a minimal two-orbital model. We label such surface modes as `shadow surface states'. 
Moreover, we show that the sufficient NNN out-of-plane hybridization leading to shadow surface states can be self-consistently stabilized for tetragonal topological Kondo insulators. 
Consequently, shadow surface states provide 
an important example of high-frequency quantum oscillations beyond the context of cubic topological Kondo insulators.
\end{abstract}

\maketitle
\section{Introduction}
%\textcolor{red}{Edge states have been the best known signature of topological phases and have been widely used as their experimental signature. Different probes such as angle resolved photo-emission spectroscopy and magnetic oscillation measurements are couple of experimental probes used to detect edge state.}  

%\textcolor{red}{Should we start with TKI?}
%\en{[EN: I think our intro is fine. We have taken sufficient pains to discon ect from SmB6 but the latter can still provide a natural intro.]}

Kondo insulators are strongly correlated systems where the hybridization between the quasi-localized $f$-electrons and the itinerant conduction electrons leads to an insulating gap
at low temperatures\cite{fiskaeppli}. 
The archetypal Kondo insulator SmB$_6$, discovered over 50 years ago\cite{Menth_PRL1968}, drew revived interest 
with proposals advancing topological surface states as an explanation for the previously-reported low- temperature resistivity plateau\cite{Dzero2010topological,Dzero2012theory,Alexandrov2013cubic,Dzero2016Top}. 
While
angle-resolved photoemmission spectroscopy (ARPES) experiments have since resolved these topological surface states\cite{Jiang_NatComm2013, Kim_NatMat2014, Neupane_NatComm2013, Frantzeskakis_PRB2013, Denlinger_review, Xu_2016} a number of puzzles remain, most notably in the linear specific heat with a large Sommerfeld coefficient\cite{Phelan_PRX2014}, gapless optical conductivity\cite{Laurita_PRB2016} and unconventional quantum oscillations\cite{Li2014two,Tan2015unconventional,Hartstein2018fermi,Hartstein2020intrinsic} (QO's).
 Although specific heat and optical conductivity anomalies 
 most likely originate from the bulk of SmB$_6$, the nature of the QO's is still under debate. Similar unconventional QO's have also been observed in another Kondo insulator YbB$_{12}$\cite{Liu2018fermi,Xiang2018quantum}, suggesting a unified underlying mechanism. \arrd{Several theoretical proposals based on magnetic breakdown\cite{Knolle2015quantum,Zhang2016quantum}, excitons\cite{Knolle_PRL2017}, impurity states\cite{Shen2018quantum,Skinner_PRM2019}, contribution from Fermi sea \cite{Pal2017quantum}, interplay between correlations and nonlocal hybridization \cite{Peters2019quantum}, oscillations of the Kondo screening \cite{Lu2020enhanced}},  fractionalization\cite{Baskaran_arxiv2015, Erten_PRL2017, Chowdhury_NatComm, Varma_arXiv2020} and surface driven mechanisms\cite{Alexandrov2015Kondo,Erten2016kondo} have been advanced in this context. 

A key aspect of QO experiments in SmB$_{6}$ is the observation of high-frequency oscillations which apparently match the extremal area of the unhybridized bulk bands.
This naturally suggests that the bulk Landau-quantized states underpin the observed oscillations.
However, ARPES experiments\cite{Jiang_NatComm2013, Kim_NatMat2014, Neupane_NatComm2013, Frantzeskakis_PRB2013, Denlinger_review, Xu_2016} on SmB$_6$ 
indicate a surface-state Fermi surface
(FS) with total area which is comparable to the extremal area of the unhybridized bands, which raises the possibility that high QO frequencies are instead rooted in the surface states.
Surface Kondo breakdown\cite{Alexandrov2015Kondo,Erten2016kondo} has been proposed as one possible mechanism for the emergence of a `large' FS  for the surface states: the Kondo effect is suppressed on the surface, liberating the conduction electrons which contribute directly to the FS of the surface states which expands as a consequence. 

Motivated by the well-studied case of SmB$_{6}$, we consider the possibility of 
high-frequency QO in a broader class of topological Kondo insulators (TKIs). In this context, we note that the areas of the surface states are not directly related with the topological nature and but depend in general on the bulk model parameters~\cite{Roy_PRB2014}. Starting from a minimal two-orbital model for cubic TKIs with nearest-neighbor (NN) hybridization, we analyze the effects of tetragonal anisotropy and next-nearest neighbor (NNN) hybridization on the surface-state spectrum. Remarkably, we find that 
the surface-states have
two FS's at charge neutrality in the extreme limit of vanishing NNN in-plane hybridization.
Due to the Kondo effect, which hybridizes quasi-localized and conduction electrons, the two bands have very different effective masses, while having equal FS areas. 
We refer to these modes as 'shadow surface states.'  The 
emergence of these shadow states stands in stark contrast to the the surface states of a 
minimal cubic model at charge neutrality, since the presence of Dirac points in that case 
effectively excludes any QO due to the surface states. Moreover, the areas enclosed by either of the two FS can be made comparable to the unhybridized area of the bulk FS. In some respect surface states are insensitive to the bulk gap and are holographic projections of bulk unhybridized bands, hence justifying the name of `shadow surface states'.
Because of the disparity in the effective masses of the 'heavy' 
hole-like and 'light' 
electron-like bands, the added contributions of the two FS in the presence of a magnetic field can give rise to high-frequency QO. 
For a finite NNN in-plane hybridization, the 
two FS's become gapped and are replaced by a set of Dirac points, with a gap controlled by the relative amplitude of in-plane and out-of plane NNN hybridization. As we discuss in greater detail below,
high-frequency QO are still possible in this case.

 %Here, we provide an alternative mechanism to obtain topological surface states with large Fermi surfaces. The key underlying feature for development of edge states with large Fermi surface is the structure of gap parameter. More specifically, we show that for a wide range of parameters for TKI, the bulk gap parameter leads to to development of edge states at the Fermi energy and with momentum parallel to the edge which is as large as the bulk Fermi momentum, when the bulk band gap is not opened. More over the decay length of edge states with large Fermi surface could be large. 
\begin{figure}
\includegraphics[width=8.5cm]{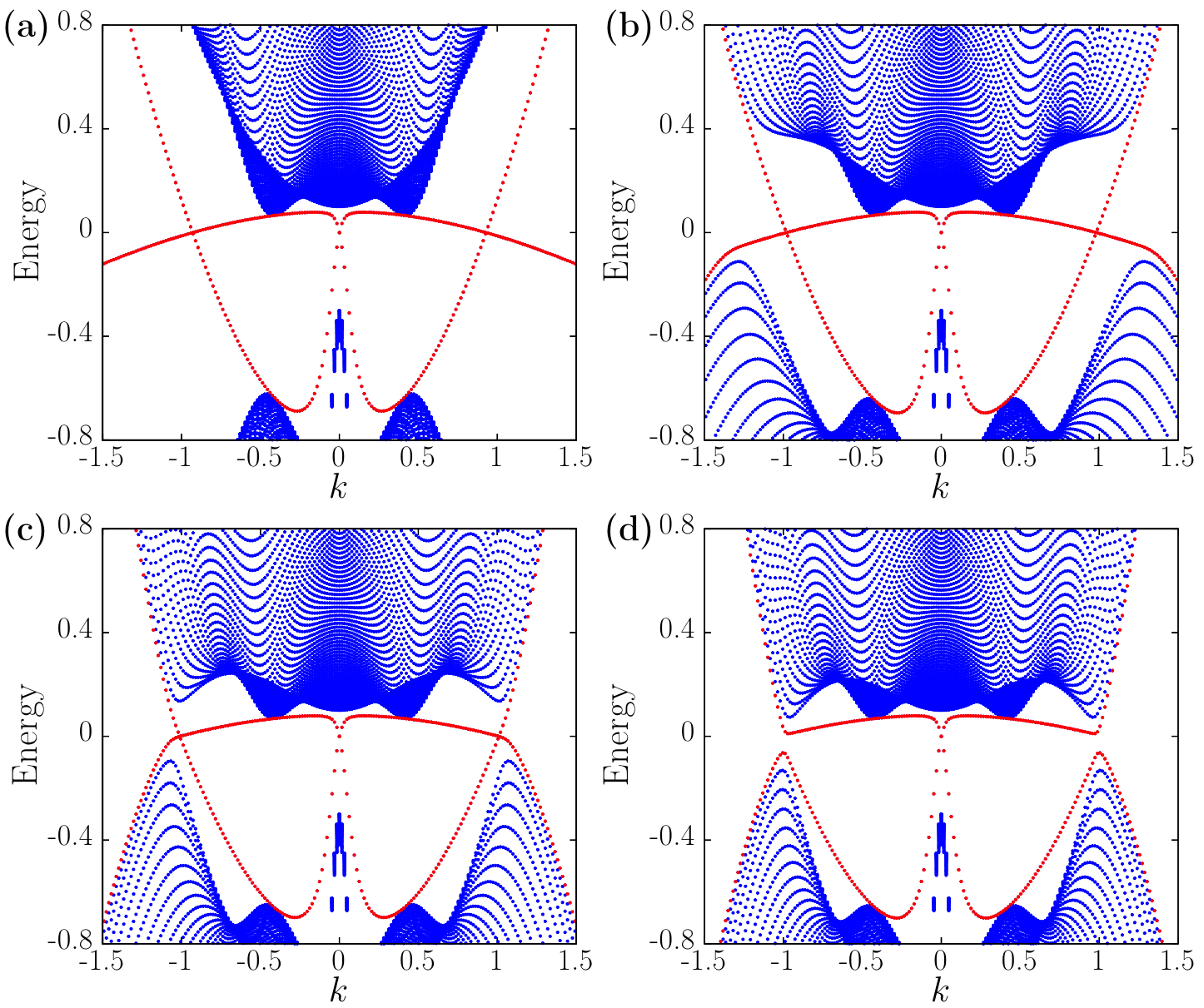}
\caption{\label{fig:EnDepkNoB} Evolution of the spectrum of a two-orbital model for topological Kondo insulators in the presence of NN and NNN hybridization terms with tetragonal symmetry. The panels correspond to solutions with increasing 
NNN in-plane hybridization, as determined by the parameter $\beta$ in (\ref{HamAddit}). 
Blue and red points denote bulk and surface states, respectively. The momentum $k$ is chosen along the $\Gamma$ to $M$ direction ($\theta_k=\pi/4$).
Panel (a) $(\beta=0)$ illustrates the emergence of shadow states. The second crossing of the two surface bands also occurs for finite momenta along an arbitrary in-plane axis, signaling the presence of a finite FS. 
Panels (b-d), corresponding to $\beta=0.2, 0.3$ and $0.35$  respectively, illustrate the presence and gapping of Dirac points. The associated Lifshitz transitions are discussed in the text. The numerical solutions were obtained for a slab geometry using a lattice model corresponding to $H_{T}$ in (\ref{HamTI}) with open boundary conditions and 200 lattice sites along the $z$ direction. The parameters of the model are $M^\prime=m^\prime=-0.45$, $M=m=-0.55$, $v=1.0$ and $\alpha=-6.0$.
}\end{figure}

In most two-orbital models of non-interacting topological insulators (TIs), the bulk band structure and gap are determined by the direct hybridization between electrons in different orbitals. By contrast, in a minimal two-orbital model for TKIs\cite{Alexandrov2015Kondo}, the effective hybridization between the orbitals, which leads to a finite gap, can be traced to the Kondo screening, at mean-field level. Consequently, the gap structure and the surface states in TKIs are not determined simply by the overlap of the weakly-interacting orbitals, but rather emerge as a consequence of Kondo interactions, and thus can exhibit a wider array of possible structures.

The  TKI mean field model, with additional NNN coupling, retains a strong topological insulator classification upon lowering the point-group symmetry from cubic to tetragonal, as indicated by the changes in the surface-state spectrum. For general NNN hybridization anisotropy, an even number of additional Dirac points appear in the surface-state spectrum. When shadow states with a finite FS emerge, the chemical potential intersects the surface-state bands at an odd number of points along any axis extending from the center of the BZ, as shown in Fig. (1) 
(a) along $\Gamma$-M.

We further show that the relative strength of NNN and NN hybridization controls the size of the shadow state FS, which approaches the extremal area of the unhybridized bands for specific value of parameters. Although a finite gap opens along the surface-state FS for any finite in-plane NNN hybridization, as previously discussed, we expect it's effect on the QO is negligible as long as the Landau level spacing is larger than the surface gap. Similar finite gap can also be opened through sharp disorder which initiates large momentum scattering. For sufficiently clean systems their effect on QO should still be minor.
Therefore, while the realized shadow states are not symmetry protected, for specific range of NN and NNN hybridizations the QO response due to the shadow states is robust.

Shadow surface states emerge in a natural generalization of a minimal model for cubic TKIs which accounts for tetragonal anisotropy and NNN hybridization. While shadow surface states are not directly relevant to cubic TKIs such as SmB$_6$ and YbB$_{12}$, they provide a striking example of surface states which can support high-frequency QO, even near the charge neutrality point in TKIs with tetragonal symmetry. In addition modification of the cubic symmetry of natural materials which accrue on the edge of the system can potentially reduce the symmetry into tetragonal structure which is required to realize the shadow states. Although such difference between the structure of the edge and bulk in some Kondo systems and its effect on their electronic properties has been discussed \cite{Poelchen2020,PhysRevB.103.165140}, realization of actual Kondo insulating material with edge or bulk properties that realize the shadow edge states is beyond the scope of this paper. We should note that recently, anomalous quantum oscillations in other materials have been observed \cite{Han2019anomalous,spl}. Given the diversity of the electronic structures and electron-electron interactions in these materials, we believe that different microscopic mechanisms might lead to these novel phenomena. On the other hand, since the appearance of shadow FS in our model relies on simple properties of effective non-interacting model of the material, we believe that our result might help to understand phenomena in a much wider class of materials.

%Consequently, they constitute a first step toward the exploration of anomalous QO beyond the context of SmB$_{6}$. 

The rest of the paper is organized as follows. 
In Sec.~\ref{ModelSec}, we introduce effective non-interacting
models for topological Kondo insulators with tetragonal symmetry and illustrate the emergence of
shadow surface states in analytical and numerical solutions of the models. In Sec.~\ref{MicroscopicModel}, 
we consider the Kondo interactions corresponding to the effective models in Sec.~\ref{ModelSec}. We show that the minimum-energy self-consistent saddle-point solutions lead to the emergence of shadow states. A detailed derivation of the analytical solutions of the the effective models introduced in Sec.~\ref{ModelSec} is presented in the Appendix.

\section{Shadow surface states}
\label{ModelSec}

The minimal, two-orbital model for cubic 3D topological insulators (TI) in the continuum limit 
has the form \cite{Qi2011Top,Zhang2009Top,Liu2010,Zhang2010First}
\begin{equation}
H_0=\left(M^\prime-m^\prime k^2\right)\tau_0+\left(M-m k^2\right)\tau_z+v\mathbf{k}\cdot\boldsymbol{\sigma}\tau_x,
\label{HamTI}
\end{equation}
where $\tau$ ($\sigma$) acts on the orbital (spin) basis. We choose units where the lattice constant $a=1$. $H_{0}$ also describes the effective mean-field Hamiltonian for 
topological Kondo Insulators (TKIs)\cite{Alexandrov2015Kondo,Dzero2016Top}.   
In most minimal models of TKIs, the narrow band originates from quasi-localized $p$ orbitals while the wide band is due to the itinerant $s$ orbital  
conduction electrons, mimicking the $f$ and $d$ orbitals in real materials. 
In contrast to conventional TIs such as Bi$_2$Se$_3$, the gap in TKIs 
emerges due to the Kondo interactions, taken at mean-field level. The nature of the insulating state 
 depends on  the sign of the product of the two mass terms with 
 $Mm>0$ ($Mm<0$) for
 strong topological or trivial 
  insulators, respectively. 
The Fermi energy of the bulk unhybridized gapless bands ($v=0$) is determined by $E_f=M'-\frac{m'}{m}M$ corresponding to the energy where the two bands cross {at momentum $k_F=\sqrt{M/m}$}.
An insulating gap appears 
as the hybridization $v$ is turned on, provided that $|m'|<m$. Furthermore, we consider cases where $Mm > 0$ such that the gapped phase corresponds to a strong TI.
The indirect gap closes when $m=m^\prime$ and for $|m^\prime|>m$ the system 
transitions to a metallic state.
We introduce boundary surfaces chosen for convenience to be perpendicular to the $z$ axis.
In the strong TI phase,
the dispersion of the surface states can be derived following the standard procedure \cite{Qi2011Top,Shan2010Eff,Shen2012book}

\begin{equation}
E^0_\pm=
E_{f} \pm v k_{||}\sqrt{1-\left(\frac{m^\prime}{m}\right)^2},
\label{EdgeTI}
\end{equation}

where $k_{||}=\sqrt{k^2_x+k^2_y}$. 
Without loss of generality, we consider the case where the chemical potential for the surface states coincides with the bulk Fermi energy $E_{f}$.
The surface bands cross the Fermi energy at $k_{||}=0$, corresponding to a 2D Dirac point.

We now consider 
additional point-group symmetry-allowed 
NNN hybridization terms 
 in the continuum limit:
\begin{align}
V_{nnn}=&\alpha v\left[k_x\left(\beta k_y^2+k_z^2\right)\sigma_x+ k_y\left(\beta k_x^2+k_z^2\right)\sigma_y \right. \nonumber \\
&\hspace{60pt} \left.+k_z\left(k_x^2+k_y^2\right)\sigma_z\right]\tau_x.
\label{HamAddit}
\end{align}

Note that the $\mathbf{k} \cdot \mathbf{\sigma}$ hybridization in (\ref{HamTI}) 
as well as $V_{nnn}$
are the limiting forms of NN and NNN hybridizations in a lattice model which preserves the $O_{h}$ cubic point-group symmetry, respectively, as shown in Sec~\ref{MicroscopicModel}.
$|\alpha|$, which was introduced in $V_{nnn}$, controls the relative strength of the NNN/NN hybridization.
Importantly, we have generalized the cubic-symmetry allowed NNN terms to allow for tetragonal  
ansiotropy corresponding to $D_{4h}$
point-group symmetry, by introducing the anisotropy factors $\beta$. Indeed, $\beta=1$ and $\beta\neq 1$ correspond to cubic and tetragonal point-groups,
 respectively. 
 Although the Hamiltonian in 
(\ref{HamTI})
 should also 
 reflect the 
tetragonal anisotropy 
 we find that neglecting this does  not qualitatively affect our results.

We now consider the Hamiltonian $H_T=H_0+V_{nnn}$, which includes the NNN hybridization terms. 
$H_T$ supports 
surface states as shown in the Appendix. To understand the structure of these states, we first consider the extreme case with vanising NNN in-plane hybridization corresponding to $\beta=0$. The surface 
 state dispersion reads:
\begin{align}
E_{\pm}=&\frac{M^\prime m^2-M m m^\prime}{m^2+\alpha^2v^2k^2_{||}}+\frac{\alpha v^2\left[m^\prime k^2_{||}+\alpha\left(M^\prime-m^\prime k^2_{||}\right)k^2_{||}\right]}{m^2+\alpha^2v^2k^2_{||}} \nonumber \\
&\pm\frac{v k_{||}\left|m+\alpha\left(M-m k^2_{||}\right)\right|}{m^2+\alpha^2v^2k^2_{||}}\sqrt{m^2-\left(m^\prime\right)^2+\alpha^2v^2k^2_{||}}.
\label{EdgeAddit}
\end{align}

%As can be seen from (\ref{EdgeTI}) the edge states are Dirac bands crossing for $k_{||}=0$ at $\mu$. 

 %Given the rotational symmetry of $H_T$ around $\hat{z}$ when $\beta=0$, without lose of generality we can drive the edge states for $k_y=0$. To see the reason why additional crossing should emerge lets for simplicity take $k_y=0$ and investigate the hybridization term of $H_T$, when $\beta=0$:
%\begin{equation}
%vk_x\left(1+\alpha k^2_z\right)\sigma_x+vk_z\left(1+\alpha k^2_x\right)\sigma_z.
%\label{JR}
%\end{equation}
%This has the form of Jackiw-Rebbi and it should support zero energy solution if $\alpha<0$, which corresponds to disappearance of hybridization term. The original crossing at $k_{1,||}=0$ is related to Jackiw-Rebbi model in $\tau$ space, whereas the new crossing at $k_{2,||}=\sqrt{M/m+1/\alpha}$ is connected to $\sigma$ space. 

%The bulk Hamiltonian with solely NN hybridization (Hamiltonian \ref{HamTI}) whose edge state given in equation (\ref{EdgeTI}) has single crossing at $k_{||}=0$. 

where $k_{\parallel}$ is the momentum in the radial direction.
In contrast to 
(\ref{HamTI}) where the two surface-state bands cross only at the Dirac point for $k_{1, \parallel}=0$, the presence of the additional bulk 
NNN hybridization in (\ref{HamAddit})
leads to additional crossings of the two surface-state bands at radial
momenta $k_{2,||}=\sqrt{M/m+1/\alpha}$. To understand why 
 the additional crossings on the circle of radius $k_{2,||}$ occur
 , we examine the real-space representation of $H_{T}$ with open boundary conditions along $z$ and $\beta=0$:

\begin{equation}
\begin{split}
H_T=&\left(M^\prime-m^\prime k_{x}^2+m^\prime\partial_z^2\right)\tau_0+\left(M-m k_{x}^2+m^2\partial_z^2\right)\tau_z \\ & + v\left[ k_{x}\left(1-\alpha \partial_z^2\right)\sigma_x-i \partial_z \left(1+\alpha k^2_x\right)\sigma_z\right]\tau_x.
\end{split}
\label{JR}
\end{equation}

%The hybridization part of the Hamiltonian (\ref{JR}) which is proportional to $v$ has the structure of the Su-Schrieffer-Heeger Hamiltonian \cite{PhysRevB.22.2099} 
%and has a zero-energy edge 
%state when $\alpha<0$. 
%Simply speaking, the zero energy edge state of the hybridization term, is effectively insensitive to the bulk gap.  

The hybridization part of the Hamiltonian (\ref{JR}) which is proportional to $v$ has the structure of the Su-Schrieffer-Heeger Hamiltonian \cite{PhysRevB.22.2099} and has a zero-energy edge 
state when $\alpha<0$ for any $k_{x}$. This edge state not being sensitive to the the presence of the bulk gap hints at the possibility that the Hamiltonian (\ref{JR}) has edge states with properties which resemble the bulk band structure when the bulk gap is closed. The condition for the presence of surface states at $E_f$ and non-zero momentum $k_{2,||}$ can be directly deduced from the full Hamiltonian $H_T$. 
We require two non-trivial eigenstates with the same sign decay lengths to satisfy the open boundary condition \cite{Qi2011Top}.
As outlined in the Appendix, the decay lengths of the two in-gap eigenstates of the Hamiltonian must satisfy
\begin{equation}
\lambda_{1} \lambda_{2}=-\frac{1}{\alpha},\end{equation}

which implies $\alpha<0$.
%\blue{[Suggestion: we can state that surface states require $\alpha<0$ and then refer to the Appendix. Alternately, we could include the Appendix here to highlight analytical solution and retain the paragraph. While useful in general, the detailed reasons for $\alpha<0$ are not really illuminating here.]}
We further require that the radial momentum of the additional crossing
$k_{2,||}=\sqrt{M/m+1/\alpha}$ 
be real which implies the following condition for the emergence of edge states with finite size FS:
 \begin{equation}
 \label{condition}\alpha<-m/M.
 \end{equation} 
In addition, with increasing $|\alpha|$, $k_{2,||}$ approaches the radius of the extremal area of the bulk unhybridized bands. 
In these cases, surface states which cross the Fermi energy at large momentum $k_{2, \parallel}$ are expected to lead to magnetic oscillations with a high frequency, which in many ways resemble the QO from the bulk Landau-levels or surface Kondo breakdown.

\begin{figure}
\includegraphics[width=\columnwidth]{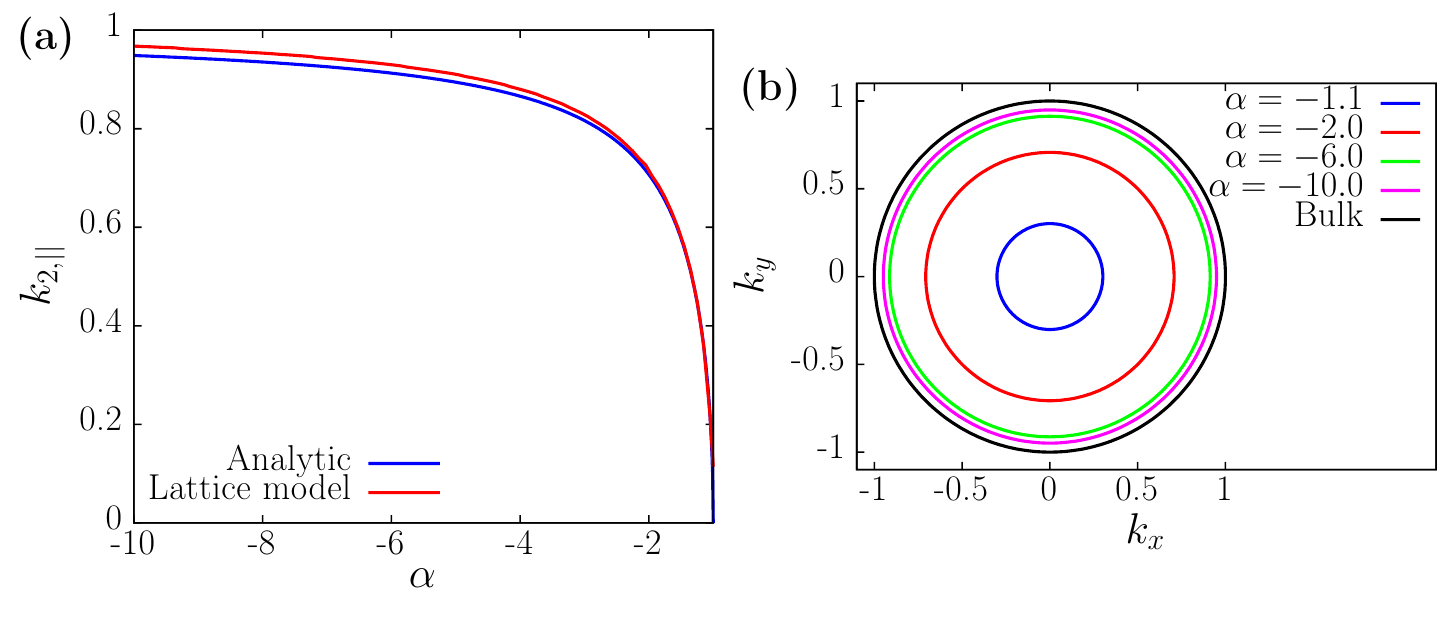}
\caption{\label{fig:FermiSurface}{(a) 
Radius of shadow state Fermi circle $k_{2,||}$ versus the NNN hybridization strength $\alpha$ as determined from the numerical and analytical solutions of $H_{T}$.
(b) Evolution of the shadow state FS with $\alpha$. The black line corresponds to the extremal-area section of the FS for the bulk unhybridized bands. For sufficiently large $\alpha$, the shadow state FS becomes comparable to the extremal area of the bulk unhybridized bands. The shadow state solutions were obtained for $\beta=0$ and all remaining parameters are the same as in Fig.~\ref{fig:EnDepkNoB}.
} 
}\end{figure}

We illustrate these arguments via a comparison between the analytical solutions of $H_{T}$ in the continuum limit and the numerical solution of a lattice tight-binding model corresponding to $H_T$ discussed in detail in the next section. The results are summarized in Fig.~\ref{fig:EnDepkNoB} (a) and Fig.~\ref{fig:FermiSurface}. Fig.~\ref{fig:EnDepkNoB} (a) shows the dependence of the lowest energy states of $H_T$ in a slab geometry with surfaces normal to the $\hat{z}$ direction. 
 
The surface states clearly show two crossings at $k_{1,||}=0$ and at finite momentum $k_{2,||}$. Fig.~\ref{fig:FermiSurface} (a) shows the dependence of $k_{2,||}$ on $\alpha$ obtained from the formula presented above and from the tight binding equivalent of $H_T$, respectively. The numerical and analytical results are in close agreement and, for $\alpha<-6$, the surface state FS approaches the extremal surface of the bulk unhybridized bands, whenever the gap vanishes, as illustrated in Fig.~\ref{fig:FermiSurface} (b).     

%To derive the condition for emergence of edge-state crossing at finite parallel momentum, penetration depth amplitudes of edge state $\lambda_1$ and $\lambda_2$, it can be shown that when $E_\pm=\mu$ and $k_{||}=k_{2,||}$ the condition $\lambda_1\lambda_2=-1/\alpha$ holds (see Appendix). In order to have finite solution at infinity $\lambda_1\lambda_2>0$ is required, which forces the same condition of $\alpha<0$ as was the case for simple Jackiw-Rebbi model. The form of the momentum $k_{2,||}$ also shows that $\alpha<-m/M$ to have real solution. 

%The presence of two crossings demonstrates that the Fermi surface at $\mu$ for $H_T$ consists of two contours. The inner contour has zero surface area, whereas the area of outer one can be controlled by the size of $\alpha$. In particular, for $\alpha\to-\infty$  $k_{2,||}{\to}k_F$. Therefore, these edge states can support large area magnetic oscillations similar to bulk bands for $v=0$ case, which we will explore below.

%\begin{figure}
%includegraphics[width=8.0cm]{FermiSurfaceNoB.pdf}
%\caption{\label{fig:FermiSurfaceNoB} Fermi contours of tight-binding equivalent of $H_T$ close to $\mu$ obtained for 50 site lattice in $z$ direction. The value of $\beta$ is $0.0$ (a), $0.02$ (b), $0.1$ (c) and $0.4$ (d). The other parameters are the same as in Fig.~\ref{fig:EnDepkNoB}.
%}\end{figure}

%Introducing $\beta$ proportional terms modify the Fermi surface. 

\begin{figure}
\includegraphics[width=8.0cm]{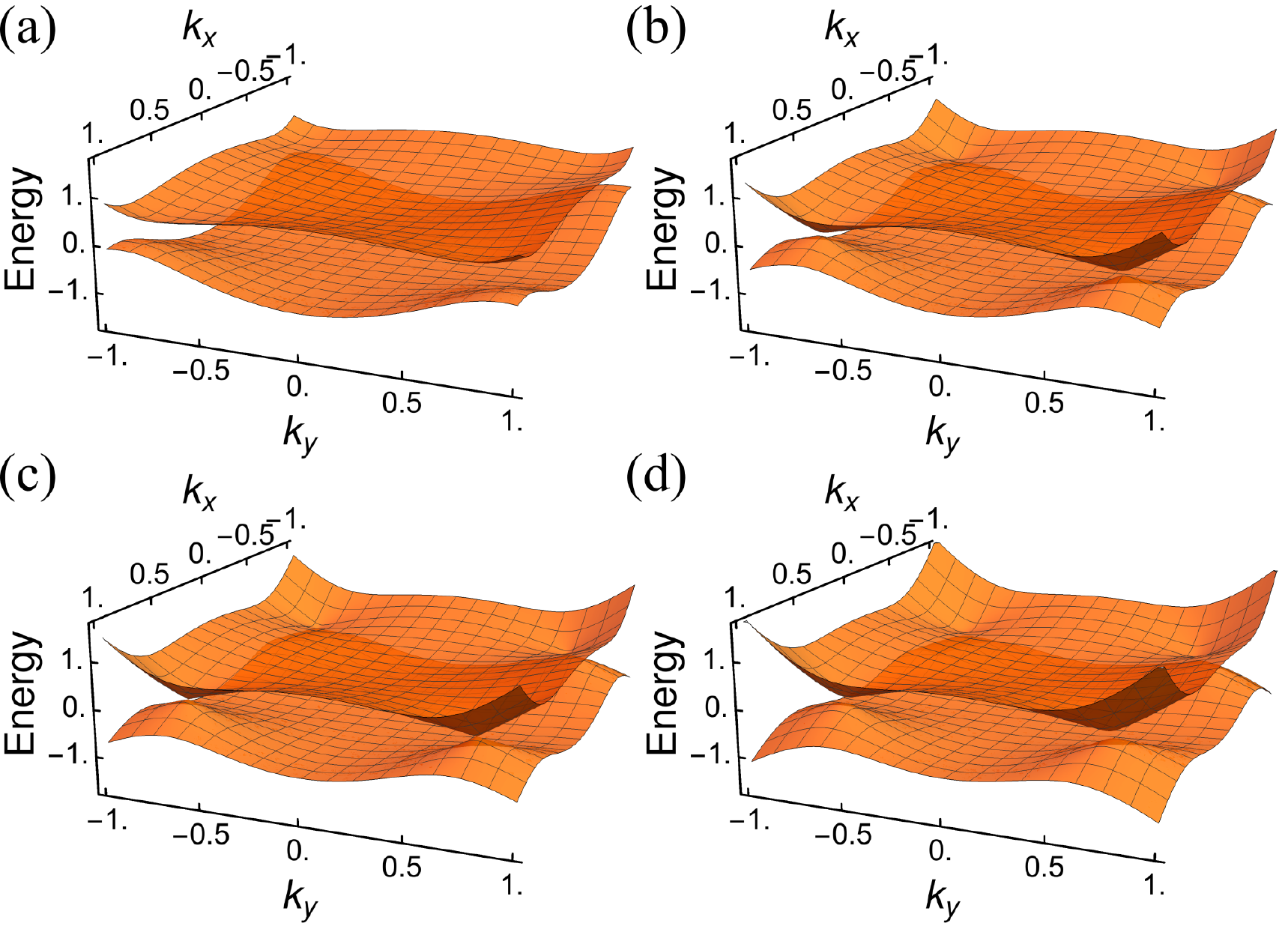}
\caption{\label{fig:BulkBands} The bulk band structure for $\beta=0.2,0.3,1/3,0.4$ (a-d) at $k_z=0$. The remaining parameters are the same as in Fig.~\ref{fig:EnDepkNoB}. For current parameters $\beta_c=1/3$. 
}\end{figure}

For $\beta=0$, the FS, which is circularly symmetric for the chosen parameters
with radius $k_{2,||}$ , corresponds to the crossing points of the two surface bands. For non-zero $\beta$,
a gap opens on the surface-state FS, except for 8 points located at $\theta_k=m\pi/2$ and $\theta_k=\pi/4+m\pi/2$ ($m$ is integer and $\tan\theta_k=k_y/k_x$). These 
correspond to surface Dirac points. We distinguish between Dirac points at $\theta_k=m\pi/2$ and $\theta_k=\pi/4+m\pi/2$. For $\theta_k=m\pi/2$ the Dirac points 
are located at $k_{2,||}=\sqrt{M/m+1/\alpha}$ (since $\beta$ dependent terms do not contribute when either $k_x$ or $k_y$ is zero), 
while for $\theta_k=\pi/4+m\pi/2$ the Dirac points occur at $k^\prime_{2,||}=\sqrt{\left(m+M\alpha\right)/m\alpha\left(1-\beta/2\right)}$. As $\beta$ increases beyond a critical value  $\beta_c=2/|\alpha| k^2_F=2m/|\alpha|M$,
 the \textit{bulk} gap closes for $\theta_k=\pi/4+m\pi/2$ and 
 the corresponding \textit{surface} Dirac points 
become 
 gapped. This is confirmed by the numerical results shown in Fig.~\ref{fig:EnDepkNoB} (b-d). 
 At $\beta=\beta_c$ the bulk band structure contains four 3D Dirac points located in the $k_z=0$ plane, along directions $\theta_k=\pi/4+m\pi/2$ and at distance $k_F$ from the center of the BZ as shown in Fig.~\ref{fig:BulkBands}. For $\beta>\beta_c$ the FS of the surface state consists of four Fermi points at directions $\theta_k=m\pi/2$.

\begin{figure}
\includegraphics[width=8.0cm]{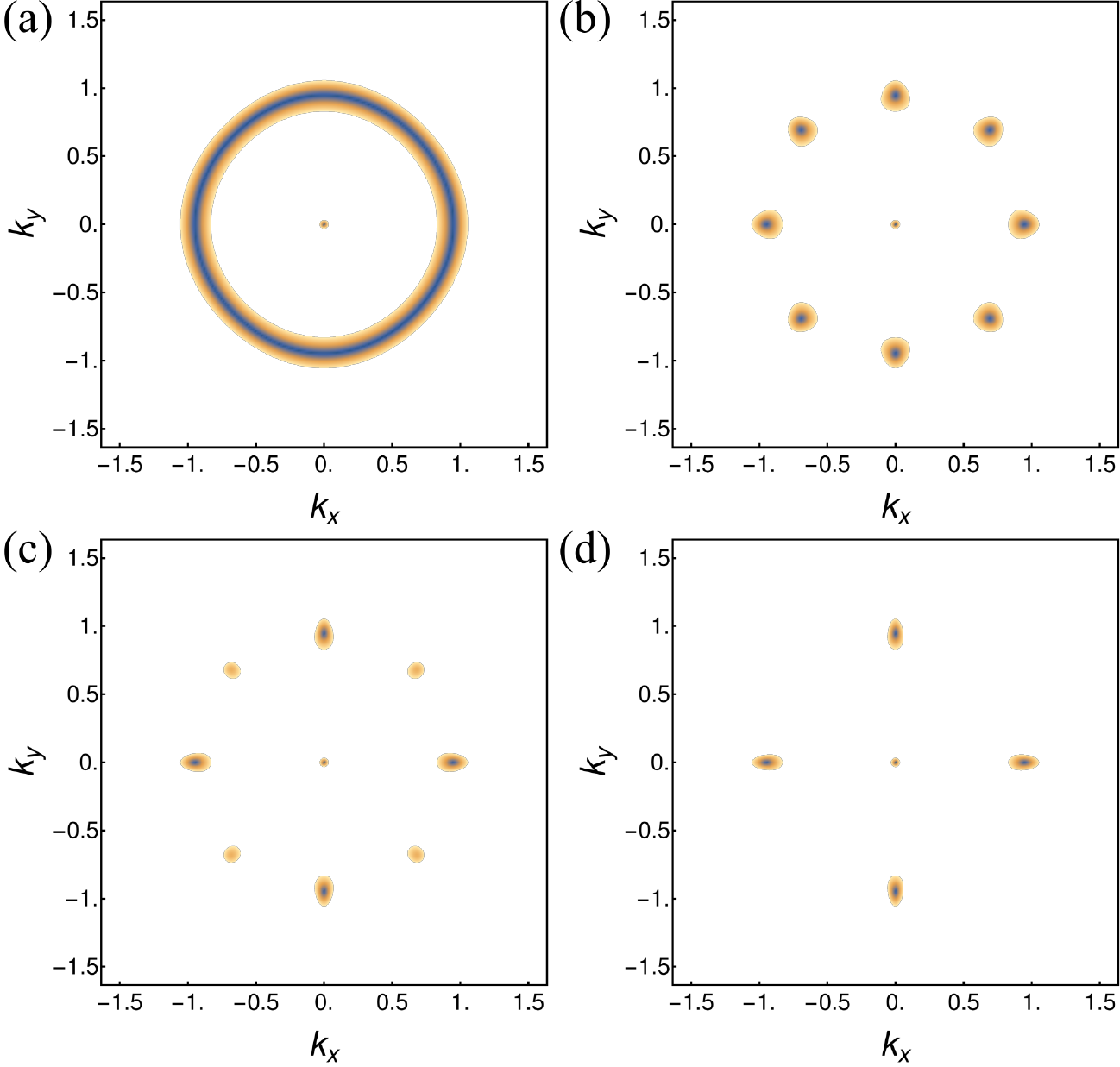}
\caption{\label{fig:FermiContour} Density plot of the surface-state gap determined from the numerical solutions for $\beta=0.0,0.2,0.3,0.35$ (a-d) respectively. The remaining parameters are the same as in Fig.~\ref{fig:EnDepkNoB}.
}\end{figure}

Fig.~\ref{fig:FermiContour} shows the evolution of 
the surface-state band gap and corresponding FS 
 extracted from the numerical solutions as a function
 of $\beta$. 
There are two Lifshitz transitions between the bands in panels (a) and (b) and (c) and (d) , respectively, as the point-group symmetry is enhanced from tetragonal to cubic by increasing $\beta$.

%The obtained results are  by numerically verified using tight-binding equivalent of $H_T$. The results are presented in Fig.~\ref{fig:EnDepkNoB}. The additional in-gap crossing of the edge states with large momentum is observed in the direction $\theta_k=m\pi/2$ for any $\beta$ and $\theta_k=\pi/4+m\pi/2$ when $\beta_c<2m/|\alpha|M$.% For larger $\beta$ that crossing is gapped out, whereas the crossing at $k_{||}=0$ stays intact.

%Fig.~\ref{fig:FermiSurfaceNoB} shows the change of the Fermi contour around chemical potential $\mu$ with the change of $\beta$. The described transitions from closed circle to eight separate small pockets is clearly observed (Fig.~\ref{fig:FermiSurfaceNoB} (a,b)). After $\beta>\beta_c$ the additional four pockets disappear as well (Fig.~\ref{fig:FermiSurfaceNoB} (d)). This shows that when $\beta$ dependent terms are present magnetic oscillations are only possible through magnetic breakdown, which is explored below.  

%Fig.~\ref{fig:FermiSurfaceNoB} shows the change of the edge-state spectrum as $\beta$ varies. At finite $\beta$ or chemical potential away from the bulk values edge fermi surfaces consist of 8 fermi pockets. 

%\section{Kondo lattice model and interaction induced shadow states}

\subsection*{Quantum oscillations induced by the shadow surface states}

\begin{figure}
\includegraphics[width=8.0cm]{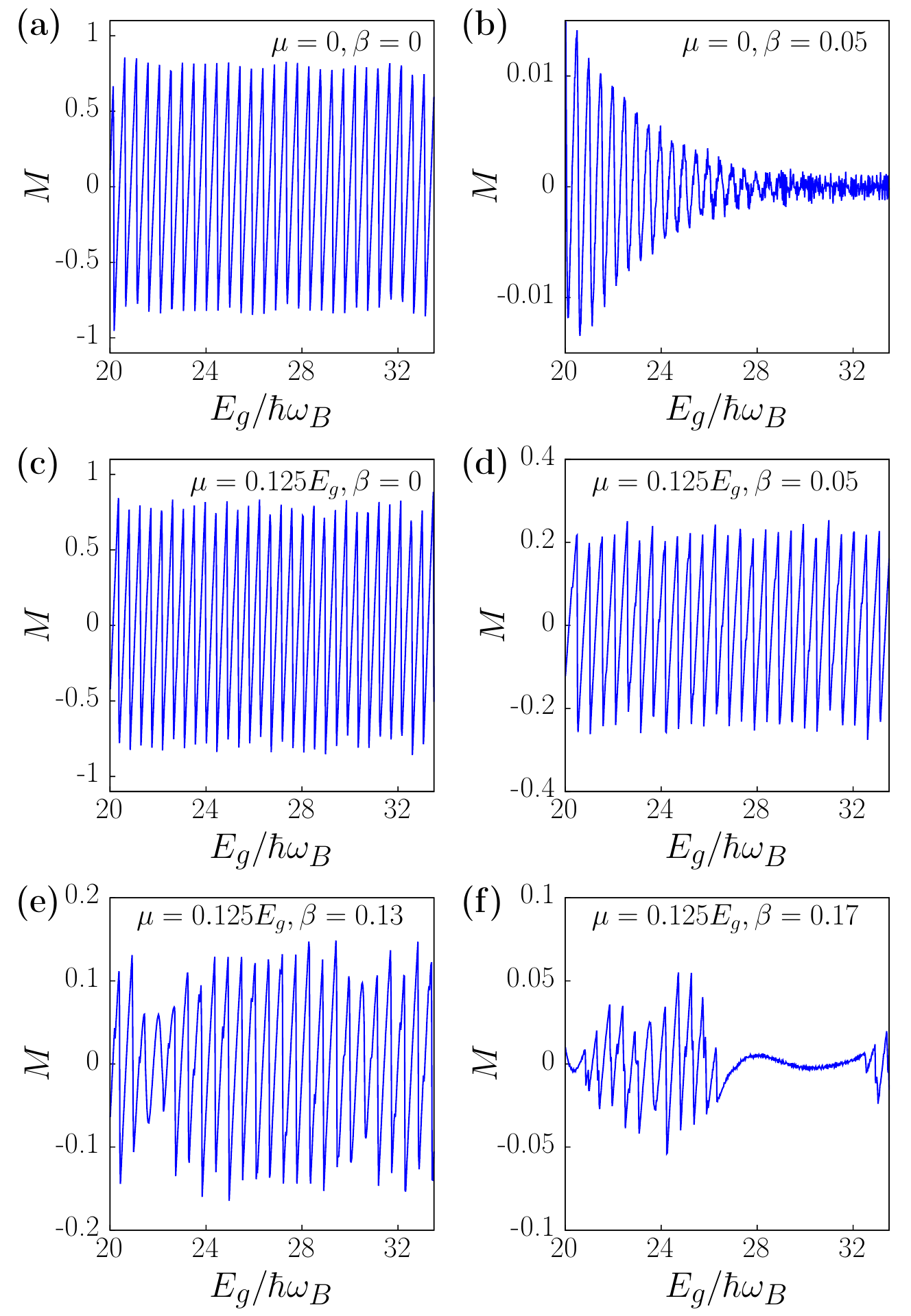}
\caption{\label{fig:Magnetization} The QO 
of the magnetization at zero temperature 
for a set of values for the chemical potential and 
anisotropy parameter $\beta$.
As discussed in the text, the ratio of the minimum direct bulk gap $E_{g}$ and the LL spacing of the unhybridized itinerant electron band $\hbar \omega_{B}$ for given magnetic field $B$ provides a measure of the contributions to the magnetic oscillations from the gapped bulk. The latter become important for $E_{g}/\hbar \omega_{B} \ll 1$. By contrast, we observe QO well above this threshold, due to the contribution of the surface states 
Note that the chemical potential is always in the bulk gap.All other parameters are the same as in Fig.~\ref{fig:EnDepkNoB}
}\end{figure}

The presence of shadow surface states with finite FS's at charge neutrality, which emerge in the extreme limit where $\beta=0$, manifests in QO of the magnetization for arbitrarily small fields. For small NNN in-plane hybridization ($\beta > 0$), the surface states become gapped except at an odd number of Dirac points. Provided that the gap is sufficiently small compared to the LL spacing, QO can still occur. Note that the surface state gap is tuned by the anisotropy parameter $\beta$ and can, in principle, be made arbitrarily small. In the most general case, QO from the surface states can thus occur for fields well below the threshold for QO from the gapped bulk. 
To show that this is the case, we calculated the magnetization of the ground state as a function of a magnetic field $B$ applied in a direction perpendicular to the surface for a set of values of the chemical potential $\mu$ and anisotropy parameter $\beta$. The results are shown in Fig.~\ref{fig:Magnetization}. Here, $E_{g}$ is the minimum value of the direct gap of the bulk states, while $\hbar \omega_{B}$ is the LL spacing for the unhybridized itinerant electron band in the presence of the field $B$. The ratio $E_{g}/\hbar \omega_{B}$ provides a threshold for QO from the gapped bulk. Indeed, this ratio decreases with increasing field and we expect QO from the gapped bulk whenever it drops below unity. By contrast, we observe QO from the surface states at ratios well above unity,  or correspondingly for fields well below the threshold value for the bulk, as illustrated in the figure.

We now turn to a detailed discussion of the results. Note that the non-oscillating part of the magnetization was subtracted. 
Panels (a) and (b) of Fig.~\ref{fig:Magnetization} 
show the QO for $\mu=0$ and for two values of the anisotropy parameter $\beta$. 
When $\beta=0$ we observe QO for all of the values of the ratio $E_{g}/\hbar \omega_{B}$, which are well away from the O(1) threshold value for QO from the gapped bulk. Note that the frequency of the oscillations is comparable to that of the unhybridized bulk bands FS for the same magnetic field B. For finite $\beta$, sharp QO are observed only for lower values of $E_{g}/\hbar \omega_{B}$ corresponding to higher fields. Due to the presence of a gap in surface state spectrum stronger magnetic fields are needed to observe quantum oscillations due to the magnetic breakdown.

%\begin{figure}
%\includegraphics[width=7cm]{MagdepTMu.pdf}
%\caption{\label{fig:QOTempDep}\arrd{Temperature %dependence of the QO amplitude $\Delta M$ for %different values of chemical potential. $\beta=0$ %and all the remaining parameters are the same as in %Fig.~\ref{fig:Magnetization}.}
%}\end{figure}
We also consider the magnetization for a finite value of $\mu$ for increasing values of $\beta$ in Fig.~\ref{fig:Magnetization} (c)-(f). 
When both $\mu$ and $\beta$ are finite, a finite FS consisting of 8 small pockets emerges, as shown in Fig.~\ref{fig:FermiContour}. The contribution of these small pockets , which oscillates with a lower frequency, becomes discernible with increasing $\beta$, as seen in panels (e) and (f).

These results illustrate that, close to the limit where the NNN in-plane hybridization vanishes, QO from the surface states with a frequency comparable to that of the bulk bands are clearly seen for magnetic fields well below the threshold for QO from the gapped bulk.

\begin{figure}
\includegraphics[width=9cm]{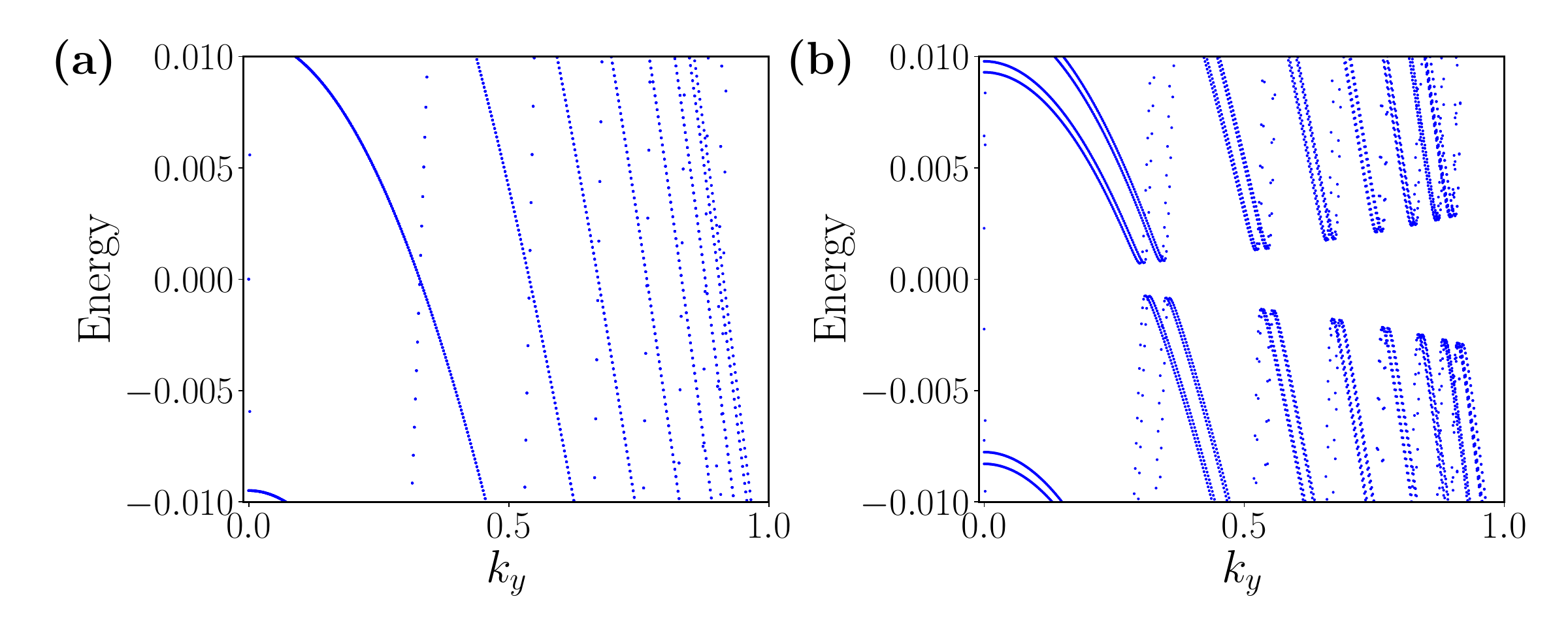}
\caption{\label{fig:EnergyTilt}Gaping of the edge states in the presence of the in plane field. (a) and (b) corresponds to $\hbar\omega_{B\parallel}=0$ and $\hbar\omega_{B\parallel}=0.025$. The results are obtained for 2D lattice in $xz$ plane with $50\times50$ sites and applying periodic boundary conditions in $x$ direction. $\beta=0$ and all the remaining parameters are the same as in Fig.~\ref{fig:EnDepkNoB}.
}\end{figure}

\arrd{One aspect of quantum oscillation, which is used to identify the two dimensional nature of conducting states, is the dependence of the frequency of oscillations on the tilt angle of the magnetic field. In conventional 2D electronic systems, the frequency of magnetic oscillations depends on the cross-sectional area of the Fermi surface normal to the field direction. This is related to the field component in the direction normal to the surface. For the shadow states, in addition to such dependence on the magnetic field direction, tilting of the field leads to gaping of the states and suppresses the magnetic oscillations. This is demonstrated numerically for the lattice model in Fig.~\ref{fig:EnergyTilt}. This effect results from hybridization of the two surface states due to the in plane component of the field and occurs at the Fermi energy where the bands cross.}  As a result, the effect of tilting of magnetic field can be used to distinguish the the surface states in this paper from other types of edge states.

\section{
Lattice models and interaction-induced shadow states}

\label{MicroscopicModel}

The emergence of shadow states in the effective continuum model requires $\alpha<-m/M$. It is crucial to examine which microscopic model could lead to effective models as in (\ref{HamTI}) and (\ref{HamAddit}) and illustrate how these microscopic models map onto the continuum limit discussed in Sec.~\ref{ModelSec}. In particular, such microscopic models involve small ratios of NN to NNN hybridizations which would be highly atypical of direct orbital overlaps in a non-interacting model. In this section we show that such hybridization structure can emerge as a result of strong interactions in TKIs.

The kinetic part of continuum model in (\ref{HamTI}) corresponds to the following form in the lattice model:
\begin{equation}\begin{split}
H_k=&\left[\left(M^\prime-6m^\prime\right)+2m^\prime\sum_i\cos\left(k_ia\right)\right]\left(f_{\textbf{k}}^\dagger f_{\textbf{k}}+c_{\textbf{k}}^\dagger c_{\textbf{k}}\right)+\\
& \left[\left(M-6m\right)+2m\sum_i\cos\left(k_ia\right)\right]\left(f_{\textbf{k}}^\dagger f_{\textbf{k}}-c_{\textbf{k}}^\dagger c_{\textbf{k}}\right),
\end{split}\end{equation}
where $f_\textbf{k}$ and $c_\textbf{k}$ are destruction operators associated with localized and itinerant bands. For simplicity we choose half-filled localized band limit. Away from $f$-electron half-filling requires mole elaborate discussion of the interacting model (see Section~\ref{KondMeanField}) which will be presented elsewhere. We should note that even though the localized $f$ band is at half-filling, the effective Heisenberg interaction between the localized moments leads to an effective dispersion for the band\cite{vojta,Alexandrov2015Kondo,Ghaemi2007higher}.

\subsection{General form of the effective hybridization in a lattice model}

We consider an effective $s-p$ orbital hybridization
 on a three-dimensional lattice in the presence of spin-orbit coupling with the most general tetragonal point-group symmetry. The form of these effective hybridization terms is determined by the point-group symmetry and does not depend on strong correlations. However, as we discuss in the next section, the effective hybridization gets renormalized due to strong correlations. 
We consider a general form which allows for tetragonal anisotropy.
The NN and NNN, symmetry-allowed hybridization terms, which include the spin-orbit coupling, in second-quantized form are\cite{Coleman2015Book}

\begin{widetext}

\begin{align}
H_{\text{Hyb}} = & \sum_{\mathbf{R}, \mu }
\bigg[
t_{0 \parallel} 
\left(
f^{\dag}_{\mathbf{R} \mu} \Psi_{0 \parallel, \mathbf{R} \mu} 
+ \text{H.c.} 
\right) 
+ t_{0 \perp} 
\left(
f^{\dag}_{\mathbf{R} \mu} \Psi_{0 \perp, \mathbf{R} \mu} 
+ \text{H.c.} 
\right)
+ t_{1 \parallel} 
\left(
f^{\dag}_{\mathbf{R} \mu} \Psi_{1 \parallel, \mathbf{R} \mu} 
+ \text{H.c.} 
\right) 
+ t_{1 \perp} 
\left(
f^{\dag}_{\mathbf{R} \mu} \Psi_{1 \perp, \mathbf{R} \mu} 
+ \text{H.c.} 
\right)
\bigg],
\end{align}

\end{widetext}

\noindent where $\mathbf{R}$ labels the sites of the cubic lattice and the $f^{\dag}_{\mathbf{R} \mu}$ operator acts on the lowest-energy Kramers doublet which emerges under the effect of the cubic or tetragonal crystal field from the $p$-orbital with spin $1/2$. Since the spin in not conserved, $\mu \in \{1,2 \}$ denotes the two components of the  Kramers doublet. We introduced the Wannier states 

\begin{align}
%%%%%%%%%%% 1st line
\Psi_{0 \parallel, \mathbf{R} \mu} = & - \frac{i}{2} \sum_{\mathbf{\hat{r}} \in \{\pm \mathbf{\hat{x}}, \pm \mathbf{\hat{y}} \} } \sum_{\nu} \left( \mathbf{\hat{r}} \cdot \mathbf{\sigma} \right)_{\mu \nu}  
c_{\mathbf{R}+ \mathbf{\hat{r}} \nu} 
\label{Eq:Wnnr_1}  \\
%%%%%%%%%%% 2nd line
\Psi_{0 \perp, \mathbf{R} \mu} = & - \frac{i}{2} \sum_{\mathbf{\hat{r}} \in \{\pm \mathbf{\hat{z}} \} } \sum_{\nu} \left( \mathbf{\hat{r}} \cdot \mathbf{\sigma} \right)_{\mu \nu}  
c_{\mathbf{R}+ \mathbf{\hat{r}} \nu} \\
%%%%%%%%%%% 3rd line
\Psi_{1 \parallel, \mathbf{R} \mu} = & - \frac{i}{2} \sum_{\mathbf{\hat{r}} \in \{\pm \mathbf{\hat{x}} \pm \mathbf{\hat{y}} \} } \sum_{\nu} \left( \mathbf{\hat{r}} \cdot \mathbf{\sigma} \right)_{\mu \nu}  
c_{\mathbf{R}+ \mathbf{\hat{r}} \nu} \\
%%%%%%%%%%% 4th line
\Psi_{1 \perp, \mathbf{R} \mu} = & - \frac{i}{2} \sum_{\mathbf{\hat{r}} \in 
\substack{ 
\{\pm  \mathbf{\hat{x}} \pm \mathbf{\hat{z}}, \\
 \pm \mathbf{\hat{y}} \pm \mathbf{\hat{z}} \}
}
} 
\sum_{\nu} \left( \mathbf{\hat{r}} \cdot \mathbf{\sigma} \right)_{\mu \nu}  
c_{\mathbf{R}+ \mathbf{\hat{r}} \nu}
\label{Eq:Wnnr_4} ,
\end{align}

\noindent defined in terms of the $s$-orbital, spin-$1/2$ conduction electrons $c_{\mathbf{R} \nu}$. By construction, each of $\Psi$'s belong to the same doublet as the $f$ fermion operators. For the most general tetragonal anisotropy, the NN hybridization splits into in-plane $t_{0 \parallel}$ and out-of-plane $t_{0 \perp}$ contributions, and similarly for the NNN terms $t_{1 \parallel}$ and $t_{1 \perp}$. Under a straightforward Fourier transformation, the hybridization terms can be re-expressed as:

\noindent \begin{align}
H_{\text{Hyb}} = & \sum_{\mathbf{k}, \mu \nu} 
\bigg[ 
V_{0, \mu \nu}(\mathbf{k}) 
+ V_{1, \mu \nu}(\mathbf{k})
\bigg] 
f^{\dag}_{\mathbf{k} \mu} c_{\mathbf{k} \nu} + \text{H.c.}
\label{Eq:Hyb}
\end{align}

\noindent where 

\noindent \begin{align}
V_{0, \mu \nu} (\mathbf{k}) = & t_{0 \parallel} \sum_{i \in \{x, y \}} s_{i} \sigma_{i, \mu \nu} 
+ t_{0 \perp} \sigma_{z, \mu \nu} s_{z} 
\label{Eq:V_0} \\
V_{1, \mu \nu} (\mathbf{k}) = & 2 \bigg[
\left(t_{1 \parallel} c_{y} + t_{1 \perp} c_{z} \right) s_{x} \sigma_{x} + 
\left( t_{1 \parallel} c_{x} + t_{1 \perp} c_{z} \right) s_{y} \sigma_{y} 
\notag \\
& + t_{1 \perp} \left( c_{x} + c_{y}\right) s_{z} \sigma_{z}
\bigg]
\label{Eq:V_1}
\end{align}

\noindent and where $s_{i} = \sin(k_{i}a), c_{i} = \cos(k_{i}a)$.

%\subsection{Effective hybridization terms for the cubic lattice} 

\subsection{
Continuum limit for cubic symmetry
}

For cases where the hybridization preserves the cubic point-group symmetry we set $t_{0 \parallel} = t_{0 \perp}$ and $t_{1 \parallel} = t_{1 \perp}$. Taking the continuum limit of the hybridization terms in (\ref{Eq:V_0}) and (\ref{Eq:V_1}) we arrive at (\ref{HamTI}) and (\ref{HamAddit}) for $\beta=1$ by identifying

\begin{align}
v&=t_0+4t_1 \\
\alpha v&=-t_1,
\end{align}
from which the condition for the emergence of shadow states,
$\alpha<-m/M$,
translates into
\begin{align}
&-4t_1<t_0<\left(\frac{M}{m}-4\right)t_1 \quad \mathrm{if} \, t_1 >0 \\
&\left(\frac{M}{m}-4\right)t_1<t_0<-4t_1 \quad \qquad \mathrm{if}  \, t_1 <0.
\end{align}

%\subsection{Effective hybridization terms for the tetragonal lattice}

\subsection{
Continuum limit for tetragonal symmetry
}

The parameters of the continuum model in (\ref{HamTI}) and (\ref{HamAddit}) are related to the lattice model parameters in (\ref{Eq:V_0}) and \ref{Eq:V_1} 
 via:
 
\begin{align}
v_\parallel&=2t_{1\parallel}+2t_{1\perp}+t_{0\parallel}, \\
v_\perp&=t_{0\perp}+4t_{1\perp}, \\
\alpha&=\frac{-t_{1\perp}}{t_{0\perp}+4t_{1\perp}}, \\
\beta&=\frac{t_{1\parallel}}{t_{1\perp}},
\end{align}

where $v_\parallel$ ($v_\perp$) is the hybridization amplitude in the 
$xy$ plane and along the $z$ direction, respectively in (\ref{HamTI}). As mentioned previously, the anisotropy of the NN
hybridization 
does not have essential consequences .

For tetragonal lattices, the condition for the surface  state crossings at finite parallel momentum (the equivalent of (\ref{condition}) when $v_\parallel\neq v_\perp$) reads as $\frac{\alpha v_\perp}{v_\parallel}<-\frac{m}{M}$. In terms of parameters of tight-binding, the required range of tight-binding model parameters read as:
\begin{equation}
\frac{-t_{1\perp}}{2t_{1\parallel}+2t_{1\perp}+t_{0\parallel}}<-\frac{m}{M}.
\label{Eq:Cndt_ttrg}
\end{equation}

For $\beta=0$ ($t_{1\parallel}=0$), this condition implies

\begin{align}
&-2t_{1\perp}<t_{0\parallel}<\left(\frac{M}{m}-2\right)t_{1\perp}\quad \mathrm{if} \, t_{1\perp} >0 \\
&\left(\frac{M}{m}-2\right)t_{1\perp}<t_{0\parallel}<-2t_{1\perp} \quad \mathrm{if} \, t_{1\perp} <0.
\end{align}

This shows that if $k^2_F=M/m<2$ the emergence of shadow states requires that $t_{0\parallel}$ and $t_{1\perp}$ have opposite signs. Recall that this conclusion is a consequence of (\ref{Eq:Cndt_ttrg}), and  is therefore rigorous only in the continuum limit when $M/m<1$. We note that for larger values of $M/m$, shadow states could develop even if $t_{0\parallel}$ and $t_{1\perp}$ have the same sign. This case corresponds to very large shadow surface state FSs and is consequently relevant in a restricted parameter-space. Therefore, we focus on $t_{0\parallel}$ and $t_{1\perp}$ having opposite sign. In Sec.\ref{KondMeanField} we demonstrate that for TKIs and due to the interacting nature of this material, such emergent pattern of parameters is energetically preferred.

\begin{figure}
\includegraphics[width=8.5cm]{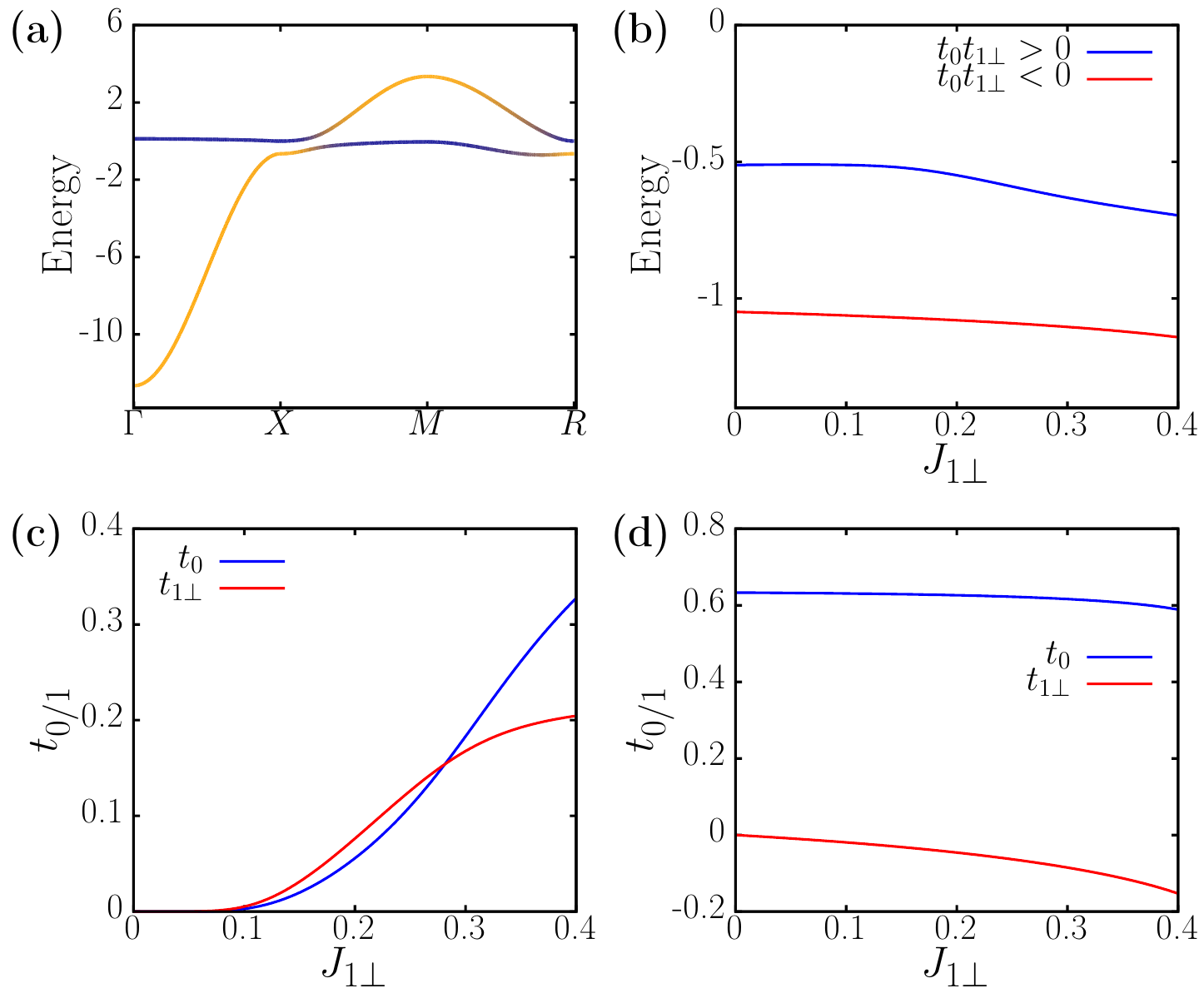}
\caption{\label{fig:SelfCons} {(a) Band structure 
determined from (\ref{Disp1}), (\ref{Disp2}) and (\ref{Eq:Hyb}) for the self-consistent solutions with effective $f$-electron level $\mu_{f}=0.66$, and NN and NNN hybridization $t_0=0.62$ and $t_1=-0.08$, respectively, which corresponds to $J_{1\perp}=0.3$. Yellow (blue) 
lines denote 
$c(f)$ character. (b) Ground-state energy per unit cell versus NNN Kondo coupling $J_{1\perp}$ for solutions where the NN and NNN effective hybridizations $t_{0}$ and $t_{1\perp}$, respectively, have equal (blue line) and opposite (red line) signs. The lowest-energy ground-state configuration occurs when the signs are opposite. Self-consistently determined NN and NNN effective hybridizations $t_{0}$ and $t_{1\perp}$, respectively, as functions of the NNN Kondo-coupling $J_{1\perp}$ for (c) $t_0t_{1\perp}>0$ and (d) $t_0t_{1\perp}<0$. The values of the parameters are: $t=1.0$, $t^\prime=0.5$, $J_{0\parallel}=J_{0\perp}=1.0$ and $J_{1\parallel}=0$, which correspond to $t_{0\parallel}=t_{0\perp}=t_0$ and $t_{1\parallel}=0$.}}
\end{figure}

%\subsection{Mean field treatment of Kondo insulators}

\subsection{
Effective hybridization from Kondo interactions
}

\label{KondMeanField}
The Kondo interaction, effectively emerges from Anderson model at half-filling  of localized fermionic band in the large interaction limit between electrons occupying the same localized orbital \cite{Coleman2015Book,Hewson}. Kondo lattice model can be derived as a lower energy effective model from periodic Anderson model using Schrieffer–Wolff transformation \cite{PhysRev.149.491}. The resulting low energy Kondo model incorporates an interaction term which is of the form of anti-ferromagnetic coupling between the spin of electrons in the local band and the spin of itinerant electrons. Given that the Kondo model is still interacting  we need to apply suitable treatment such as mean-field approximation\cite{Coleman2015Book,Ghaemi2007higher}. Such treatments were used to capture phenomena such as large FS and heavy quasiparticle mass in heavy-Fermi liquids\cite{senthil2003, senthil2004}. In this section, we show that $t_{0\parallel}$ and $t_{1\perp}$ having opposite signs is energetically preferred at mean-field level for TKIs. In the half-filling limit, we consider the NN and NNN Kondo interaction\cite{Dzero2016, Alexandrov_PRB2014, Ghaemi2007higher,Ghaemi2008angle}

\begin{align}
H_{K} = &  \sum_{\mathbf{R}} \bigg[ J_{0 \parallel} \mathbf{S}_{0 \parallel, \mathbf{R}} \cdot \mathbf{S}_{f, \mathbf{R}} 
+  J_{0 \perp} \mathbf{S}_{0 \perp, \mathbf{R}} \cdot \mathbf{S}_{f, \mathbf{R}} 
\notag \\
& + J_{1 \parallel} \mathbf{S}_{1 \parallel, \mathbf{R}} \cdot \mathbf{S}_{f, \mathbf{R}} 
+  J_{1 \perp} \mathbf{S}_{1 \perp, \mathbf{R}} \cdot \mathbf{S}_{f, \mathbf{R}} 
\bigg]
\end{align}

\noindent where 

\begin{align}
\mathbf{S}_{f, \mathbf{R}} = & \frac{1}{2} \sum_{\mu \nu} f^{\dag}_{\mathbf{R} \mu} 
\mathbf{\sigma}_{\mu \nu} f_{\mathbf{R} \nu}. 
\end{align}

\noindent For tetragonal symmetry, the in-plane and out-of-plane Kondo couplings are different with $J_{0/1 \parallel} \neq J_{0/1 \perp}$. Note that the $f$ fermion is constrained to the half-filled Fock space. Similarly, the operators 

\noindent \begin{align}
%%%%%%%%%%% 1st line
\mathbf{S}_{0 \parallel, \mathbf{R}} = & \frac{1}{2} \sum_{\mu \nu}  \Psi^{\dag}_{0\parallel, \mathbf{R}  \mu} \mathbf{\sigma}_{\mu \nu}  \Psi_{0\parallel, \mathbf{R}  \nu}
\\
%%%%%%%%%%% 2nd line
\mathbf{S}_{0 \perp, \mathbf{R}} = & \frac{1}{2} \sum_{\mu \nu}  \Psi^{\dag}_{0\perp, \mathbf{R} \mu} \mathbf{\sigma}_{\mu \nu}  \Psi_{0\perp, \mathbf{R} \nu}
\\
%%%%%%%%%%% 3rd line
\mathbf{S}_{1 \parallel, \mathbf{R}} = & \frac{1}{2} \sum_{\mu \nu}  \Psi^{\dag}_{1\parallel, \mathbf{R} \mu} \mathbf{\sigma}_{\mu \nu}  \Psi_{1\parallel, \mathbf{R}  \nu} \\
%%%%%%%%%%% 4th line
\mathbf{S}_{1 \perp, \mathbf{R}} = & \frac{1}{2} \sum_{\mu \nu}  \Psi^{\dag}_{1\perp, \mathbf{R} \mu} \mathbf{\sigma}_{\mu \nu}  \Psi_{1\perp, \mathbf{R} \nu}
\end{align}

\noindent are defined in terms of the Wannier states in (\ref{Eq:Wnnr_1})-(\ref{Eq:Wnnr_4}).

A straightforward decoupling of the Kondo interactions in the particle-hole channel reproduces (\ref{Eq:Hyb}) if we identify

\begin{align}
\label{SelfCons1}
t_{0 \parallel} = & - \frac{3J_{0 \parallel}}{8} \bigg \langle\sum_{\mu} \Psi^{\dag}_{0 \parallel, \mathbf{R} \mu} f_{\mathbf{R} \mu} \bigg \rangle 
\\
t_{0 \perp} = & - \frac{3J_{0 \perp}}{8} \bigg \langle\sum_{\mu} \Psi^{\dag}_{0 \perp, \mathbf{R} \mu} f_{\mathbf{R} \mu} \bigg \rangle \\
t_{1 \parallel} = & - \frac{3J_{1 \parallel}}{8} \bigg \langle\sum_{\mu} \Psi^{\dag}_{1 \parallel, \mathbf{R} \mu} f_{\mathbf{R} \mu} \bigg \rangle 
\\
t_{1 \perp} = & - \frac{3J_{1 \perp}}{8} \bigg \langle\sum_{\mu} \Psi^{\dag}_{1 \perp, \mathbf{R} \mu} f_{\mathbf{R} \mu} \bigg \rangle
\label{SelfCons4}
\end{align}

We choose a tight-binding part for conduction and $f$ electrons \cite{Alexandrov2015Kondo}:
 
\begin{align}
\label{Disp1}
\epsilon_\mathbf{k}&=-2t\sum_i\cos k_i-4t^\prime\sum_{i\neq j}\cos k_i\cos k_j,\\
\epsilon_\mathbf{f k}&=-\eta\epsilon_\mathbf{k}+\mu_f,
\label{Disp2}
\end{align}

where $\eta=0.01$ 
and $\mu_f$ is the chemical potential for the 
$f$ electrons. We impose half-filling for both f and conduction electrons:
\begin{align}
1&=\frac{1}{N_k}\langle\sum_{\mathbf{k} \mu}f^\dagger_{\mathbf{k}\mu}f_{\mathbf{k}\mu}\rangle, \\
1&=\frac{1}{N_k}\langle\sum_{\mathbf{k}\mu}c^\dagger_{\mathbf{k}\mu}c_{\mathbf{k}\mu}\rangle,
\end{align}
where $N_k$ is the number of points in the bulk BZ.
Finally, both $t_{i \parallel / \perp}$ and $\mu_{f}$ are determined self-consistently. 
%{\blue{Note that our effectively non-interacting model is not the same as a periodic Anderson model which has strong interactions on the f-electrons. However, our model can be derived from a periodic Anderson model in two steps: first a Schrieffer-Wolff transformation to obtain the corresponding Kondo model, followed by a mean field treatment of the Kondo model. The Kondo model is particularly relevant for half-filled localized band in strong interaction limit \cite{Hewson,Coleman2015Book}.}}

Fig.~\ref{fig:SelfCons} (a) shows the band structure of the Hamiltonian formed from (\ref{Disp1}), (\ref{Disp2}) and (\ref{Eq:Hyb}). 
The ground state has band inversion at the 3 $M$ points in the in the {BZ} and therefore has a nontrivial topological Z$_{2}$ index.
Fig.~\ref{fig:SelfCons} (b-d) shows the solution of (\ref{SelfCons1}-\ref{SelfCons4}) for half-filling, when $t_0=t_{0\parallel}=t_{0\perp}$ and $t_{1\parallel}=0$. We distinguish solutions with $t_0$ and $t_{1\perp}$ having the same and opposite sign. We find that the self-consistent solutions with $t_0 t_{1\perp}<0$ minimize the ground-state energy whereas solutions with $t_0 t_{1\perp}>0$ are always higher in energy as shown in Fig.~\ref{fig:SelfCons} (b). Fig.~\ref{fig:SelfCons} (c) and (d) show the corresponding $t_{0/1\perp}$ as a function of $J_1$ for $J_0$ = 1. At $J_1=0$ the lower energy solution corresponds to finite $t_0$ and $t_{1\perp}=0$. Moving away to finite $J_1$ the $t_0 t_{1\perp}<0$ solution smoothly connects with that low energy solution, whereas $t_0 t_{1\perp}>0$ solution develops in different region and is never preferred.

%\blue{REMOVE(Areg):Fig.~\ref{fig:SelfCons} (b) shows the total energy of the system per unit cell. Solutions where $t_0$ and $t_{1\perp}$ have opposite signs always minimize the ground-state energy per unit cell, as shown in  Fig.~\ref{fig:SelfCons} (b).} %As discussed previously, shadow states emerge in this regime. While the orbital diagonal terms in (\ref{Disp1}) and (\ref{Disp2}) are different from those in Sec.~\ref{ModelSec}, shadow states still emerge for a subset of the parameter values in Fig.~\ref{fig:SelfCons} (d).

%Fig.~\ref{fig:SelfCons} (b-d) shows the solution of (\ref{SelfCons}) for half-filling. As can be seen from the figure the amplitude of $b$ is enhanced when $t_0$ and $t_1$ have opposite signs, indicating a stronger Kondo-induced hybridization. The regime of opposite sign NN and NNN hybridization terms.  As shown previously the surface states have non-trivial dispersion and additional crossings for finite momentum. Although we considered a cubic point-group symmetry in this section, tetragonal anisotropy introduced via a finite $\beta$ parameter introduced in Sec.~\ref{ModelSec} can also be included in the current model. Our results remain valid for that case as well. 

\section{Conclusions}

Motivated by the well-studied case of SmB$_{6}$, we considered a natural extension of a minimal two-orbital model with NN hybridization for cubic TKIs to cases with tetragonal anisotropy and additional NNN hybridization. In the tetragonal limit with strictly NNN out-of-plane hybridization, we found that the surface states show two FS's with different effective masses at the charge neutrality point. 
The existence of two FS's, one hole-like and another  electron-like, with equal areas but very different effective masses is inherent to TKI's, due to the hybridization between quasi-localized and itinerant electrons. These features, in turn make finite-frequency QO possible. 
This contrasts with the more common situation in TI's, which typically exhibit Dirac points at charge neutrality and no QO from the surface states for vanishing magnetic fields.
The QO 
due to the surface states can also occur in presence of finite NNN in-plane hybridization or disorder, provided that the gaps induced by these mechanisms are comparable to the Landau level spacing.

In addition, we showed that a predominantly NNN out-of-plane hybridization realizes the minimum-energy ground-state configuration in TKIs with tetragonal lattice symmetry. 
Consequently, shadow surface states provide a concrete example of how arbitrarily high-frequency QO can occur beyond the immediate context of cubic TKIs such as SmB$_{6}$. 

Finally, we note that the ideal candidates for the emergence of shadow surface states with a small gap are compounds with a dominant NNN out-of plane hybridization. The most natural realizations are therefore topological Kondo insulators under strong compressive strain.

\section*{acknowledgement} PG acknowledges support from National Science Foundation Awards No. DMR-1824265 for this work. A.G.~acknowledges support from the European Union's Horizon 2020 research and innovation program under the Marie Sk\l{}odowska-Curie grant agreement No. 754411. EMN is supported by ASU startup grant. OE is in part supported by NSF-DMR-1904716.

\section{Appendix}
\label{Append}

%\subsection{A}
In this appendix, we discuss the analytical solutions of the surface states of continuum-limit Hamiltonian $H_T=H_0+V_{nnn}$ introduced in Sec.~\ref{ModelSec}. Taking into account the double-degeneracy of the states  due to inversion and time-reversal symmetry, we consider the following two degenerate eigenstates ansatz for the general solutions of $H_{T}$ (in the basis $\left\{|1,\uparrow\rangle,|2,\uparrow\rangle,|1,\downarrow\rangle,|2,\downarrow\rangle\right\}$, $\{1,2\}$ denoting orbital index and $\{\uparrow,\downarrow\}$ denoting the spin):

\begin{widetext}
\begin{equation}
\psi^1_{\xi}(z)=N_1\left(\begin{array}{c}
L_-+m_-\lambda^2_\xi-E \\
iv\lambda_\xi\left(1+\alpha k^2\right) \\
0 \\
-vk_x\left(1+\alpha\left(\beta k^2_y-\lambda^2_\xi\right)\right)-ivk_y\left(1+\alpha\left(\beta k^2_x-\lambda^2_\xi\right)\right)
\end{array}\right)e^{-\lambda_\xi z},
\end{equation}
\begin{equation}
\psi^2_{\xi}(z)=N_2\left(\begin{array}{c}
-vk_x\left(1+\alpha\left(\beta k^2_y-\lambda^2_\xi\right)\right)+ivk_y\left(1+\alpha\left(\beta k^2_x-\lambda^2_\xi\right)\right)\\
0 \\
-iv\lambda_\xi\left(1+\alpha k^2\right) \\
L_++m_+\lambda^2_\xi-E
\end{array}\right)e^{-\lambda_\xi z},
\end{equation}
\end{widetext}
where $L_\pm=M_\pm-m_\pm k^2_\parallel$, $M_\pm=M^\prime\pm M$, $m_\pm=m^\prime\pm m$, $\xi=1,2$ and $\lambda_\xi$ are determined from a characteristic equation\cite{Shan2010Eff,Shen2012book}. We focus on non-trivial surface states of a semi-infinite system with an open boundary condition at $z=0$ and normalization condition of the wave function in the region $z>0$. Therefore, we are only interested in $\lambda_\xi$ with positive real part. Using the definitions

\begin{align}
F&=m_+\left(L_{-}-E\right)+m_-\left(L_{+}-E\right)\nonumber \\
&+v^2\left(1+4\alpha k_\parallel^2+\alpha^2k_\parallel^4+4\alpha^2\beta k^2_xk^2_y\right) \\
R&=F^2-4\left(m_{+}m_{-}-\alpha^2v^2k_\parallel^2\right)\times \nonumber \\
&\left[\left(L_{+}-E\right)\left(L_{-}-E\right)-v^2\left(k_\parallel^2+\alpha\beta k^2_x k^2_y\left(4+\alpha\beta k_\parallel^2\right)\right)\right]
\end{align}

we express the solutions via 

\begin{align}
\lambda_\xi&=\left[2\left(m_{+}m_{-}-\alpha^2v^2k_\parallel^2\right)\right]^{-\frac12}\times \nonumber \\
&\hspace{60pt}\left(-F+(-1)^\xi\sqrt{R}\right)^{\frac12}  
\end{align}
Non-trivial solution satisfying boundary condition at $z=0$ are obtained from the requirement $\det\{\psi^1_{1}(0),\psi^1_{2}(0),\psi^2_{1}(0),\psi^2_{2}(0)\}=0$, which takes simple form at
%By imposing an open boundary condition at $z=0$, we  obtain the condition for the emergence of topological surface states, which takes simple form at 
$k_{2,||}$ when $\beta=0$ or $\theta_k=m\pi/2$: 
\begin{equation}
\lambda_1\lambda_2=-\frac{1}{\alpha},
\end{equation}
For $\beta\neq0$, and $\theta_k=\pi/4+m\pi/2$ also for
$k^\prime_{2,||}$, the condition is
\begin{equation}
 \lambda_1\lambda_2=\frac{1}{\beta-2}\left(\frac{M\beta}{m}+\frac{2}{\alpha}\right).
 \label{BetanzCond}
\end{equation}
Non-trivial surface states thus require $\lambda_1\lambda_2>0$ and since $0<\beta<1$, the above conditions can only be satisfied if $\alpha<0$. In addition, the condition (\ref{BetanzCond}) requires $\alpha>-2m/M\beta$ which is equivalent to $\beta<\beta_c$ derived in main text.

For general  momentum $k_{||}$, the non-trivial surface state spectrum is
\begin{widetext}
\begin{align}
E_\pm=&\frac{M^\prime m^2-M m m^\prime}{m^2+\alpha^2v^2k_\parallel^2}+\frac{\alpha v^2\left(m^\prime k_\parallel^2+\alpha\left(M^\prime-m^\prime k_\parallel^2\right)k_\parallel^2+2\alpha\beta m^\prime k^2_x k^2_y\right) }{m^2+\alpha^2v^2k_\parallel^2}\pm \nonumber \\
&\frac{v}{m^2+\alpha^2v^2k_\parallel^2}\left[\left(m^2-m^{\prime 2}+\alpha^2v^2k_\parallel^2\right)\left(\left(m+\alpha\left(M-m k_\parallel^2\right)\right)^2k_\parallel^2+\alpha^2\beta^2m^2k^2_x k^2_y k_\parallel^2+\right.\right.
\nonumber \\
&\left.\left. 4\alpha\beta m k^2_x k^2_y\left(m+\alpha\left(M-m k_\parallel^2\right)\right)+\alpha^4 \beta^2 v^2 k^2_x k^2_y\left(k^2_x-k^2_y\right)^2\right)\right]^{1/2}.
\end{align}
\end{widetext}
Above spectrum reduces to the from given in (\ref{EdgeAddit}) for $\beta=0$.

\bibliography{KondoTI}
\bibliographystyle{apsrev4-1}

\end{document}